\documentclass[a4paper,11pt]{article}

\usepackage{graphicx}
\usepackage{amsmath}
\usepackage{amsthm}
\usepackage{amssymb}
\usepackage{pifont}

\usepackage{multirow}
\usepackage[left=20mm,right=20mm,top=20mm,bottom=20mm]{geometry}
\small \normalsize

\begin{document}

\begin{center}

  {\LARGE \bf 
Is neuroscience facing up to statistical power?}

\vspace{3mm}

{\large Geoffrey J. Goodhill}

\vspace{1mm}

Queensland Brain Institute and School of Mathematics and Physics, \\The
University of Queensland, St Lucia, QLD 4072, Australia.\\
Email: g.goodhill@uq.edu.au

\vspace{1mm}

{\em \today}

\end{center}




\section*{Abstract}

It has been demonstrated that the statistical power of many
neuroscience studies is very low, so that the results are unlikely to
be robustly reproducible. How are neuroscientists and the journals in
which they publish responding to this problem? Here I review the
sample size justifications provided for all 15 papers published in one
recent issue of the leading journal {\em Nature Neuroscience}. Of
these, only one claimed it was adequately powered. The others mostly
appealed to the sample sizes used in earlier studies, despite a lack of
evidence that these earlier studies were adequately powered.  Thus,
concerns regarding statistical power in neuroscience have mostly not
yet been addressed.

\section*{Introduction}

It is well-documented that the biomedical sciences are beset by bad
statistical practices, and that this is one of the reasons for the
current `reproducibility crisis' \cite{ioannidis05,colquhoun14}.
Prominent amongst these problems are n values that provide only
low statistical power. Genuine effects that do actually exist are
missed, and many effects that are found to be significant are likely
to be just random chance. Neuroscience is no exception to this rule
\cite{button13}. Indeed, due to the particular challenges of this
field, studies which can be completed within traditional parameters of
time, cost and ethical approval are often restricted to low n values.
Furthermore this data is then analysed post-hoc from many different
perspectives in the hope of finding significant results, further
increasing the probability of false positives.

The purpose of the present article is not to review again these
problems, which are well documented. Rather, I consider how the
community of authors, reviewers and journal editors in neuroscience is
responding to this clearly visible challenge. The leading neuroscience
journal {\em Nature Neuroscience} provides a good opportunity to do
this, since (unlike most neuroscience journals) it requires authors to
provide answers to some basic statistical questions about
the design of their experiments. It is therefore of interest to see
what answers have been forthcoming in recently published
papers. Presumably, because the papers were published, the authors,
reviewers and editors all thought these answers were acceptable. 

Here I reproduce the statements regarding sample size from all 15
papers published in the August 2016 issue, and find that all of them
except one essentially confess they are probably statistically
underpowered. I do not explicitly identify which papers these came
from, because my goal is not to cast doubt on any specific work: this
is simply a (somewhat) random subset to illustrate a very broad
issue. Furthermore, there is no reason to think these problems are any
different in other journals (though a recent study has argued that
statistical power is {\em negatively} correlated with journal impact
factor \cite{szucs16}).  What makes {\em Nature Neuroscience}
attractive for analysis in this regard, besides its current ranking as
the highest-impact primary research journal in the field, is that it
takes the trouble to require authors to make explicit comments about
certain statistical matters.

\section*{Statements of powerlessness}

These are ordered thematically, and do not reflect the ordering within the issue.

1. {\em The sample size for each experiment was determined based on
  published studies using similar experimental designs together with
  pilot experiments from our laboratory. This allowed us to determine
  the sample size required for each experiment to ensure a statistical
  power of 0.8 and an alpha level of 0.05.}

Here the authors clearly address the issue of statistical
power. Although potentially one might want to see the evidence for the
claim, this statement provides reassurance that these
results are likely to be reproducible.

2. {\em Sample size was predetermined on the basis of published
  studies, experimental pilots and in-house expertise.}

It is encouraging that pilot studies were undertaken, but it is
unclear how these pilots or the in-house expertise were used to
determine statistical power.

I now group several statements together, since they are all very similar.

3. {\em Sample sizes for each condition of this study are similar to
  those generally employed in the field\ldots and were not
  predetermined by a sample size calculation.}

4. {\em No statistical methods were used to predetermine sample sizes,
  but our sample sizes are similar to those generally employed in the
  field.}

5. {\em Sample size choice was based on previous studies, not
  predetermined by a statistical method.}

6. {\em No statistical tests were used to predetermine sample sizes, but
  our sample sizes are similar to those in previous studies}

7. {\em No statistical methods were used to predetermine sample sizes,
  but our sample sizes are similar to those generally employed in the
  field.}

8. {\em Group sample sizes were chosen on the basis of previous studies.}

9. {\em No statistical methods were used to predetermine sample sizes,
  but our sample sizes are similar to those previously reported.}

10. {\em No statistical methods were used to pre-determine sample sizes
  but our sample sizes are larger to those reported in previous
  publications.}

The obvious problem with all these statements is that they do not
address whether any of these previous studies demonstrated they were
adequately powered (that previous work produced significant results
says nothing about statistical power, a basic point that appears not
to be widely appreciated).  In addition, unless exactly the same
experiments were performed, the variability and effects sizes are
likely to be different, meaning that the sample size required to
achieve adequate power will also be different.

11. {\em No statistical methods were used to predetermine sample sizes,
  but the tissues were randomly chosen in each age group\ldots and
  uniformly processed. Also, our samples sizes are similar to those of
  the discovery set of a similar experimental design in a previous
  publication}

Besides the problems mentioned above, this statement conflates statistical
power with other issues of experimental design.

12. {\em No statistical methods were used to predetermine sample
  sizes. Sample size was decided on the basis of our previous
  experience in the field and was not pre-determined by a sample size
  calculation. The sample size are similar to those generally employed
  in the field and is justified by the high rate of exclusion due to
  the difficulty of the combined methodological approaches}

Here the authors appeal to practical limitations on sample
sizes. These limitations are real and worthy of
acknowledgement. However this does not provide information
pertinent to the statistical power, and thus reproducibility, of the
results.

13. {\em No estimates of statistical power were performed before
  experiments; animal numbers were minimized to conform to ethical
  guidelines while accurately measuring parameters of animal
  physiology.}

Here the authors appeal to ethical limitations on sample sizes. Again,
while real and worthy of acknowledgement, the same arguments apply as
mentioned above.

Finally, we come to  perhaps the two most worrying statements. 

14. {\em No statistical tests were used to predetermine sample sizes, but
  our sample sizes are similar to those generally employed in the
  field. Normal distribution of data was assumed, but not formally
  tested.}

15. {\em Normality of the data distributions was assumed, but not
  formally tested.}

The last makes no statement about sample sizes at all, despite this
supposedly being a requirement of the journal. More importantly, both
statements explicitly state that the authors do not know whether the
statistical tests they applied were actually appropriate.

For comparison the statements in the July 2016 issue were very
similar. One article justified sample sizes in terms of a power
calculation, while the remainder bar one (which simply stated that `no
statistical methods were used to predetermine our sample sizes')
appealed to similarity with sample sizes used in previous studies, in
one case in a different species.

\section*{Discussion}

All of the statements reviewed above were approved by the authors,
reviewers, and journal editors, and one must therefore conclude that
they reflect currently accepted practice in the field. It is widely
known and understood that statistical power is a key issue affecting
reproducibility, yet 14/15 of these statements (93\%) do not address
statistical power. Most of them appeal to precedent for sample sizes,
despite the facts that most neuroscience studies are underpowered
\cite{button13}, and that new experiments will most likely have
different variances and effect sizes from previous work. It is clearly
a step forward that {\em Nature Neuroscience} requires authors to
explicitly answer some key questions regarding statistical
analysis. However, that the journal is willing to accept answers which
are clearly inadequate, and even sometimes admissions that the
statistical analysis performed was quite possibly wrong, suggests that
the journal is still contributing to the problem rather than the
solution. For comparison the relatively new journal {\em eNeuro}
requires authors to provide the statistical power of each experiment
reported. However this is merely the observed power, which provides
little or no additional information beyond the observed p value
\cite{lenth01}, and thus this policy does not help matters much
either.

I am not attempting to single out these 14 papers as being of any more
concern than any other work in the field, rather they simply provide a
revealing window on community standards at the highest
level. Neuroscientists seem willing to accept that work in the field
generally uses low n values. Sometimes this might be reasonable: for
instance the effect size of the difference in phenotype between a
wild-type and knockout may be very large (though even in
cases such as these power calculations are rarely provided). Certainly
many important findings have been robustly reproduced, even though the
statistical power of the original (or indeed subsequent) results was
not established (e.g. \cite{bliss73}).  However, in many experiments
the effects are subtle, and low n values and thus power mean that, on
average, reproducibility will be low.

This is a very difficult problem (which I hasten to add I am also
struggling with in my own research).  Neuroscience experiments are
often intrinsically long-term and low-throughput. For instance
uncovering the function of a disease-related gene in a mouse model, or
studying the neural correlates of consciousness in an awake behaving primate
model, can require large resources and many years of work to obtain a
single main result. Many of the latest techniques are extremely
technically challenging, and therefore (as alluded to in one of the
statements above) a large proportion of experiments fail. This can
lead to a big mismatch between the n values at the start and end of
the experiment. Funding is tight, and increasing n by even one animal
for a particular experiment can have costly implications in time and
money.  Increasing ethical pressures on the use of animals in research
(as alluded to above) add additional constraints on the n values that
are practically achievable. However there is clearly also a cultural
component to sample sizes in neuroscience: for instance work in
organisms such as {\em C. elegans} and {\em Drosophila}, to which some of the
above constraints are less applicable, also do not usually consider statistical power.

What can be done?  Clearly better education for neuroscientists (at
all levels) regarding statistical issues is important to address the
lack of statistical scepticism that apparently plagues the field, and
books such as \cite{reinhardt15} should be more widely read and
understood (for instance many neuroscientists still appear to think
that obtaining $p < 0.05$ means it is 95\% likely they have discovered
something true, no matter how small the n value \cite{halsey15}).
From the perspective of a field such as statistical genetics, one
solution seems obvious: neuroscientists should collaborate in larger
teams and share data, so that many small and weakly powered results
can be replaced with a few strongly powered results. However, while
data sharing in general should and is being broadly encouraged, there
are problems with this general model for the community of
neuroscientists at large (for an interesting discussion see
\cite{mainen16}). How would everyone agree what were the right
experiments to do?  How would cutting-edge and often highly
non-trivial methodologies be standardized between labs?  How would
this model not be detrimental to the entreprenurial spirit that has
fuelled so many important discoveries in neuroscience?  The risk is
that progress in neuroscience, where publication rates are already low
compared to some other fields of biomedical science (for the reasons
mentioned above), could be reduced to an unviable level.

Another approach is to replace the ubiquitous current statistical
paradigm, of null hypothesis significance testing, with estimation
\cite{cumming17}. This can be done by confidence intervals and/or
Bayesian approaches. Now, instead of there being a `bright line of
truth' at an arbitrarily chosen probability threshold relating to the
null hypothesis (which, after all, is not what one is actually
interested in), the focus is on determining degrees of confidence in
quantities such as the effect size and the difference between
means. Binary statements about `significance' versus `nonsignificance'
are replaced with graded confidence variations which are easier to
interpret intuitively. However, this has yet to catch on in
neuroscience.

Staying within the confines of null hypothesis significance testing, I
have argued that neuroscientists seem willing so far to accept the
current situation regarding (lack of) statistical power. Perhaps that
is indeed the best that can presently be achieved, given the current
statistical paradigm and practical constraints in the field. Perhaps
we should just accept that most studies will be likely underpowered
and reproducibility will likely be a recurring issue, at least until
some more mature stage of development is reached. However if this is
the case, it would be helpful if authors, reviewers and journal
editors more clearly acknowledged that underpowered studies lead to
weak and potentially irreproducible results.

\end{document}